\def\year{2017}\relax
\documentclass[letterpaper]{article} 
\usepackage{DataforPolicy}  
\usepackage{times}  
\usepackage{helvet}  
\usepackage{courier}  
\usepackage{url}  
\usepackage{graphicx}  
\usepackage{bm}  
\usepackage[utf8]{inputenc}
\usepackage{fourier} 
\usepackage{array}
\usepackage{makecell}
\usepackage{dblfloatfix}
\begin{document}
%
\title{Evaluating the impact of the 2012 Olympic Games policy on the regeneration of East London using spatio-temporal big data }

\author{Xiao Zhou\\
Computer Laboratory\\
Univerisity of Cambridge\\
xz331@cam.ac.uk
\And
Desislava Hristova\\
Computer Laboratory\\
Univerisity of Cambridge\\
desislava.hristova@cl.cam.ac.uk
\And
Anastasios Noulas\\
Data Science Institute\\
University of Lancaster\\
a.noulas@lancaster.ac.uk
\And
Cecilia Mascolo\\
Computer Laboratory\\
Univerisity of Cambridge\\
cecilia.mascolo@cl.cam.ac.uk
}

\maketitle
\begin{abstract}
For urban governments, introducing policies has long been adopted as a main approach to instigate regeneration processes, and to promote social mixing and vitality within the city. However, due to the absence of large fine-grained datasets, the effects of these policies have been historically hard to evaluate. In this research, we illustrate how a combination of large-scale datasets, the Index of Deprivation and Foursquare data (an online geo-social network service) could be used to investigate the impact of the 2012 Olympic Games on the regeneration of East London neighbourhoods. We study and quantify both the physical and socio-economic aspects of this, where our empirical findings suggest that the target areas did indeed undergo regeneration after the Olympic project in some ways. In general, the growth rate of Foursquare venue density in Olympic host boroughs is higher than the city’s average level since the preparation period of the Games and up to two years after the event. Furthermore, the deprivation levels in East London boroughs also saw improvements in various aspects after the Olympic Games. One negative outcome we notice is that the housing affordability becomes even more of an issue in East London areas with the regeneration gradually unfolding.  
\end{abstract}

\section{Introduction}
In recent decades, the topic of urban regeneration has received increasing attention globally, due to the undeniably significant role it may play in city development (Ye, et al., 2015). It is used by governments as a boost to disinvested urban areas, reducing relative levels of deprivation and promoting ‘social mixing’ (Keddie, 2014). Furthermore, for international metropolises like London, urban regeneration can be even more necessary by building attractive urban spaces of business, leisure and residence to 
promote its functional role in the global economy (Sassen, 2001; Webber, 2007).  

Realising the positive effects that might be brought with regeneration, the urban government of London has put forward a series of policies to instigate it, so as to counterbalance the continuing post-war exodus from the city centre and draw developers to invest (Keddie, 2014). Among them, the regeneration policies related to 2012 Olympic Games are generally regarded as the most striking ones recently. “The Game will be London’s single most important regeneration project for the next 25 years”, as the Mayor of London indicated. This mega event was expected to originally stimulate the regeneration agenda of six host boroughs (Hackney, Tower Hamlets, Newham, Barking and Dagenham, Waltham Forest, and Greenwich) and eventually catalyse the revitalisation of the whole East London. Compared with the rest of the city and the country, East London is one of the most physically fragmented, environmentally compromised and socially deprived districts which is seen as the greatest remaining regeneration opportunity in inner London (ODA, 2010).  

However, the success of regeneration through Olympic Games is not guaranteed. With quite a few challenges lying ahead, long-term efforts and study are required. Firstly, the deprivation of East London is a historical problem that is extremely hard to solve. Even though several waves of regeneration strategies have already been implemented in East London, tackling the deprivation of this area is still challenging for many decades (Gavin, 2012; Dmitry, 2012). Some critics suggest that the Olympic strategy might be going in the same direction as previous policies in its failure to achieve the regeneration goal (NEF, 2008). Furthermore, the experience of previous Olympic cities shows that the mega events are not necessarily an effective mechanism for catalysing urban development and renewal (Dmitry, 2012). Some Olympic projects turned out to be ‘white elephants’ only in the years after the event (Galina, 2011; Sue, 2010; Cashman, 2009; Macrury, 2008). Also, to evaluate the impact of regeneration policy is hard in nature due to the lack of data and data-driven approaches. Indeed, from existing evidence, the power of the state to intervene and influence the market through making policies has been demonstrated. However, what to and how to measure the physical and social regeneration that is induced by policies is still a new area of inquiry for researchers to work out (Lees \& Melhuish, 2015). Whether or not Olympic regeneration project can be shown to have long-term benefits is still unclear in the existing literature. 

Fortunately, with the emergence of the Big Data era, an increasingly rich data environment has become available to policy makers to shed new light on the issue. Through the lens of a huge range of data sources, the change of urban places and actual behaviours of citizens have become unprecedentedly observable at very fine grain spatial and temporal scale, offering the potential to benefit urban study. Some promising results have been presented in existing research using Foursquare data. For instance, Daggitt et al. (2015) analysed urban growth across major cities worldwide. Karamshuk et al. (2013) studied optimal retail store placement in the city. In this research, we use the English Indices of Deprivation and Foursquare data to examine how successfully the Olympic regeneration policies have worked and question whether the ideals of promoting local liveability and social-economic development have been achieved in East London at different geographic levels. In order to answer this overarching question on whether the policies are effective, the research is structured around the three sub-questions: 1. What the representative Olympic regeneration policies and their goals are; 2. Whether the policies have brought significant changes to various categories of urban places within the target area; 3. How human spatio-temporal mobility patterns have been influenced since the policies took place; 4. And whether the deprivation conditions of East London become more or less divergent. 

\section{Policies}
From the very beginning, the ambition of bridging the historical gap between West and East London has been at the core of the Olympic bid and frequently shown in the subsequent related plans (Dmitry, 2012). These plans are led by governments and organisations whose interests range from comprehensive coverage of the Games’ impact to a focus on specific geographical and subject areas (DCLG, 2015). In some Olympic legacy plans at region or even country level, the regeneration of East London was put forward as one of the major targets. A summary of the key plans and their goals involved in the renewal of East London are provided in Table 1. Some other policies make their targets more specific by focussing on the regeneration of smaller spatial coverage, like the host boroughs or the Olympic Park. The Strategic Regeneration Framework (SRF) is one of the significant plans that contributes to the delivery of this aim and is the main policy that is evaluated in this research.

\begin{table}[h!]
	\normalsize
	\begin{tabular}{|c|c|}
		\hline
		\textbf{Plans} & \textbf{Targets}\\
		\hline
		London Plan (2004) & Regenerating East London \\
		\hline
		\makecell{Mayor of London’s five\\ legacy promises (2007)} &  \makecell{•Transforming the \\heart of East London}\\
		\hline
		\makecell{The Government’s \\ Legacy Action \\ Plan (2008)} & \makecell{•Transforming the heart\\ of East London\\•Making the Olymlic Park\\ a blueprint for \\sustainable living} \\
		\hline
		\makecell{The Coalition\\ Agreement (2010) } & \makecell{• Ensuring that the Olympic \\ Park can be developed after\\ the Games as one of the\\ principal drivers of \\regeneration in East London.} \\
		\hline
		\makecell{Department for \\Communities and \\Local Government’s \\(DCLG) Business Plan \\2011-2015 (2010)} & \makecell{•Regenerating the \\Thames Gateway } \\
		\hline
	\end{tabular}
	\caption{Groups of London Wards in ANOVA analyses.}
\end{table}

The objective of the SRF is to ensure both of the physical regeneration and socio-economic benefits are brought to the host boroughs.  From the physical regeneration side, it aims to create well-designed, successful and sustainable places that attract new business, new mixed communities, and enhance existing neighbourhoods. On the socio-economic part, it holds an ambitious goal to achieve radical socio-economic convergence between host boroughs and the London average for key indicators of deprivation. To meet the convergence objective, seven core outcomes were proposed in the SRF (Table 2).  

The SRF expects to achieve the goals within a 20-year timeframe from 2010 to 2030. It also includes three interim targets to be met by 2015, which are: narrowing the gap of local satisfaction between London average and the host boroughs; delivering new and better places to live and work, including homes, the related schools, health centres and other social infrastructure; and completing the early stages of the Olympic Park redevelopment as a lasting legacy (SRF, 2009). 

\begin{table}[h!]
\centering
\normalsize
	\begin{tabular}{|c|c|}
		\hline
		1 & \makecell{Creating a coherent and high quality\\city within a world city region}\\
		\hline
		2 & \makecell{Improving educational attainment,\\ skills and raising aspirations}\\
		\hline
		3 & \makecell{Reducing worklessness, benefit\\ dependency and child poverty} \\
		\hline
		4 & Homes for all\\
		\hline
		5 & Enhancing health and wellbeing \\
		\hline
		6 & \makecell{Reducing serious crime rates\\and anti-social behaviour} \\
		\hline
		7 & \makecell{Maximising the sports legacy\\and increasing participation} \\
		\hline
	\end{tabular}
	\caption{Seven key outcomes of the SRF}
\end{table}

\section{Impact Evaluation}
In this section, the datasets collected for the research are described, followed by how they are analysed to evaluate the physical and socio-economic impact of the Olympic Games on East London (primarily focus on six Olympic host boroughs in this study) at different geographical levels. 

\subsection{Data Collection}

Data for the study are drawn from the English Index of Deprivation and location-based social network Foursquare. 

The English Indices of Deprivation is an official measure of relative deprivation for small areas (Lower-layer Super Output Areas) in England calculated by the Department for Communities and Local Government. It is organised across seven sub domains of deprivation (Health Deprivation and Disability; Employment Deprivation; Income Derivation; Education, Skills and Training Deprivation; Crime; Barriers to Housing and Services; and Living Environment Deprivation) to produce the Index of Multiple Deprivation (IMD), an overall relative measure of deprivation. The versions used in this research are 2010 Index and 2015 Index, through a comparative analysis of which, changes of multiple deprivation levels of the study area are tracked to have an insight of the socio-economic impact of the policy. Alongside the Index of Deprivation data, we also collected user mobility records and venues' information in London through a four-year long dataset from June 2010 to August 2014 of Foursquare. This Foursquare data contains all "transitions" (pairs of check-ins by users between two different venues) occurring within London during the study period. For each transition, we have the start time, end time, source venue and destination venue. The information of Foursquare venues in London is also available in our dataset, through which we know the geographic coordinates, category, creation time and the total number of users that have check-in(s) at each venue.

\subsection{Physical Impact}
To explore whether the policy has significantly influenced the physical regeneration in Olympic host boroughs, we use Foursquare data to identify local urban growth patterns by measuring whether there are many venues of certain categories emerging after the policy was implemented. Figure 1 shows the location of our research area and the distribution of venues at the end of the study period. In the following analysis, we track the trend of venue growth and category composition by time and discuss how the changes vary by host boroughs using London as benchmark comparator. 

\begin{figure}
	\centering
	\includegraphics{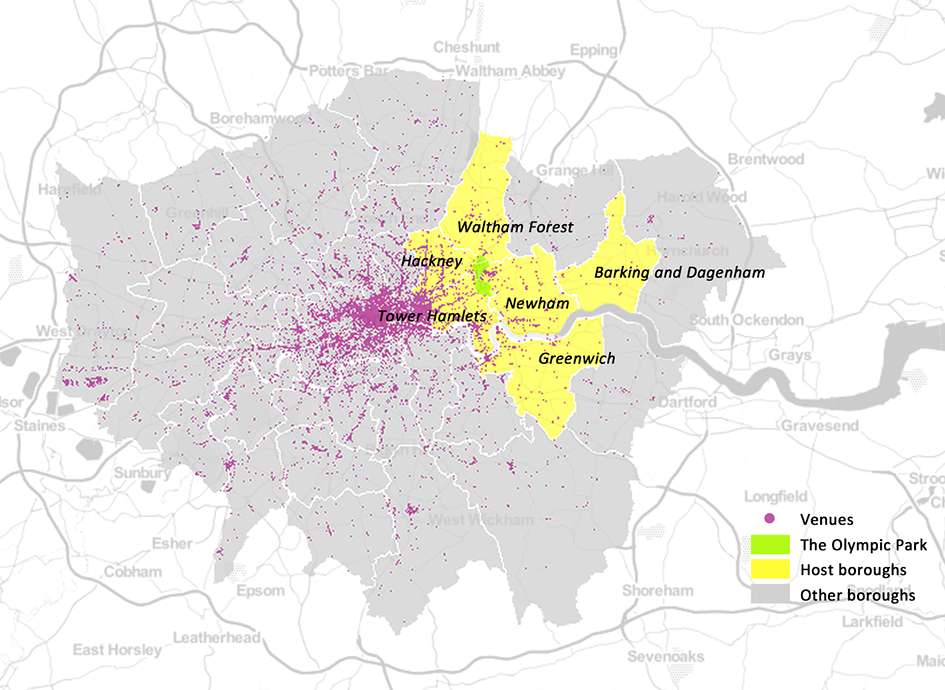}
	\caption{The distribution of the Olympic Park, Host boroughs, and Foursquare venues by the end of the study period}
\end{figure}

\subsubsection{Venue Density}
We explore how the density of venues changed within host   boroughs in total, and compare it with the average level in the whole city. The venue density is calculated by dividing the number of venues within an area by its size. As can be seen from the first subfigure of Figure 2, both of the densities in host boroughs and London present a growing trend during the four years. By the end of the study period, the venue density of host boroughs in total was almost twice that of its initial value. One noteworthy point is the relative change between the two densities. At the beginning, the venue density of host boroughs was lower than London’s average level. Then the gap became increasingly smaller until July 2013, when the values became equal. After that, the average venue density of host boroughs overpassed the city average level by the end of August 2014. In the middle subfigure of Figure 2, the monthly growth rates of venue densities are presented, from which we can see host boroughs have a higher growth rate in general than London’s, and two evident peaks show in October 2011 and August 2012, respectively. Instead of studying the host boroughs as a whole, the analysis next is to look into the differences on venues’ growth between boroughs. As the last subfigure of Figure 2 presents, Hackney and Tower Hamlets are the boroughs where the venue densities are far more than the average and other host boroughs. Another striking point is the venue density boost of Newham (where the Olympic Park is located) in October 2011.

\begin{figure*}
	\centering
	\includegraphics{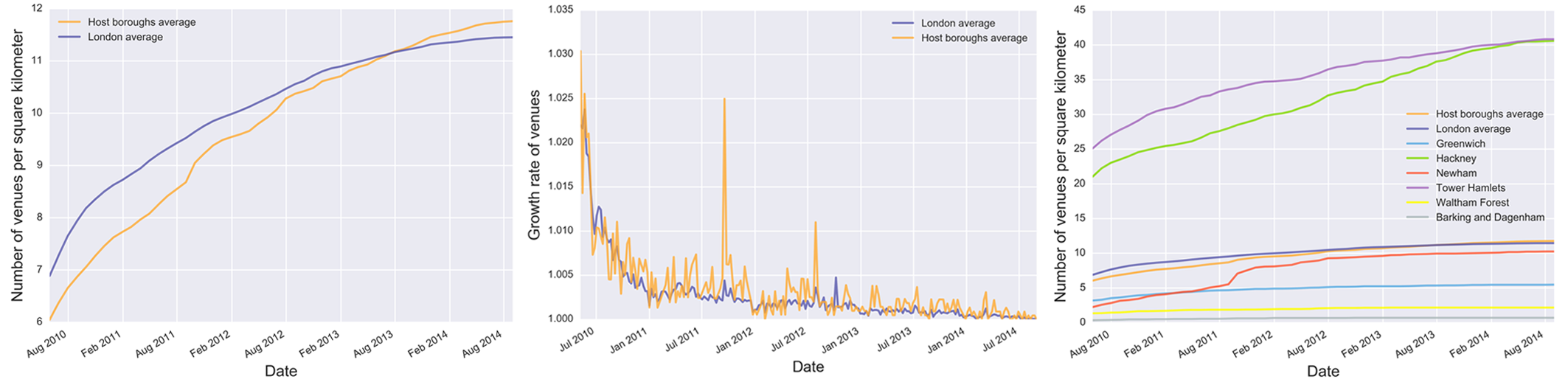}
	\caption{The comparison of venue density between host boroughs and London average by date }
\end{figure*}

\subsubsection{Venue Composition}

After understanding how the density of venues has changed, we explore the difference in composition of new venues between a host area and its counterparts. In Figure 3, the colour of each square represents the proportion of venues of that particular category from the total number of new venues within the area in each year. We observe that for both host boroughs and the rest of London, the category of food has the largest percentage among each year’s new venues. However, the proportion of food for host boroughs was not as significant as for the rest of London in general. An important component of new venues in host boroughs is the Outdoors \& Recreation category, which always presents the second largest category within the group, and is usually above the proportion for the rest of London. This phenomenon is especially evident during the years of 2010, 2012, and 2014, when the proportion of host boroughs as compared to the rest are 10\% to 6\%, 6.5\% to 3.4\%, and 5.9\% to 0.76\%, respectively. Another four categories that the Olympic area shows a greater proportion for are Arts \& Entertainment, Professional \& Other Places, Nightlife Spots, and Stadiums.  On the contrary, categories related to transport, like Airports, Train Stations, Travel \& Transport are more expressed in the rest of London. 

\begin{figure*}[b!]
	\centering
	\includegraphics{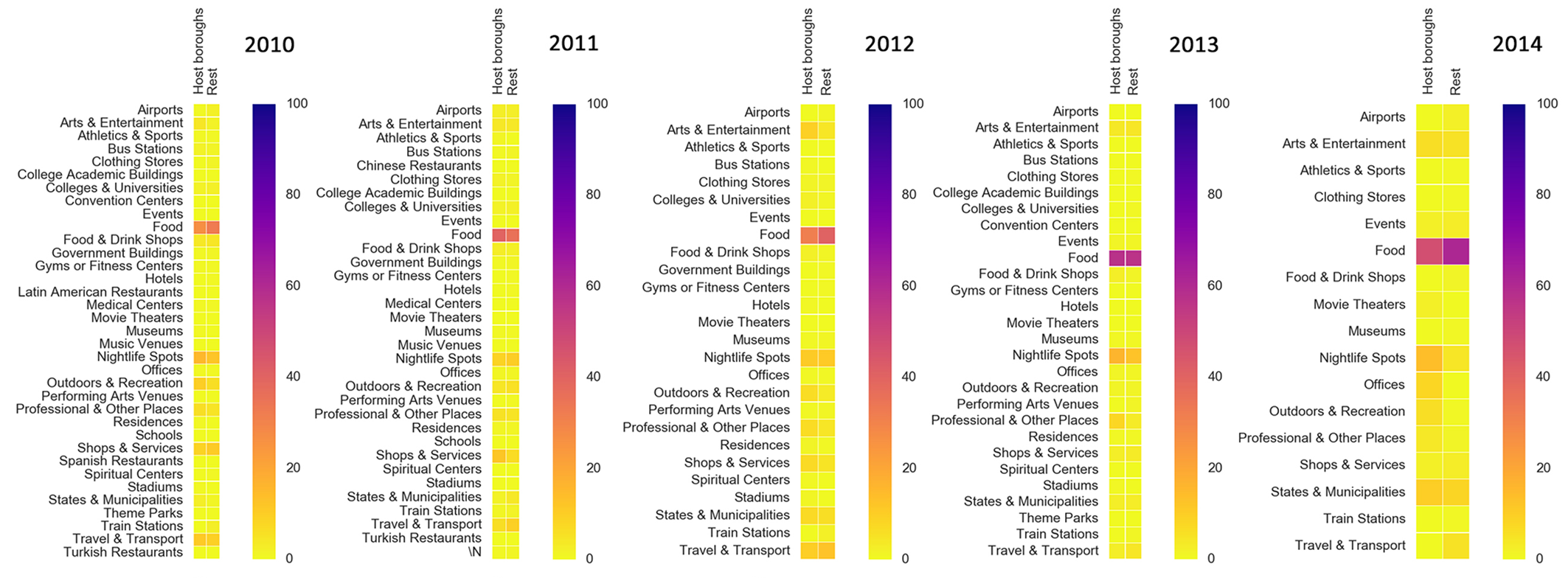}
	\caption{The percentage of categories for new venues in each year}
\end{figure*}

\subsection{Socio-economic Impact}
Compared with physical legacies, the ‘soft’ impacts of socio-economic are less apparent, harder to measure, but necessary in the evaluation of Olympic policy’s effectiveness. In this section, we try to measure the socio-economic impact through studying human mobility features, and an investigation of deprivation changes before and after the Olympic regeneration program.  

\subsubsection{Mobility Features}

To explore how the volume of transitions from and towards an area changes over time may provide us some alternative insights into local vitality. We analyse this in Figure 4 and Figure 5, which display the transition popularities in host boroughs and London in total. We measure the transition popularity through dividing the total number of transitions from and towards an area by its population. We observe that the Olympic host boroughs as a whole has lower number of flows than that in London in most cases, with an exception in August 2012, when the Olympic Games took place. This phenomenon is more evident when the analysis is undertaken at borough’s level. As we can see from Figure 5, in the major Olympic borough of Newham, the flow transitions peaked during the Olympic period with an incomparable value. However, the average number of transition flows in host boroughs fell dramatically after the mega event, and returned to its normal level in the post-Olympic period.  

\begin{figure}
	\centering
	\includegraphics[width=8cm]{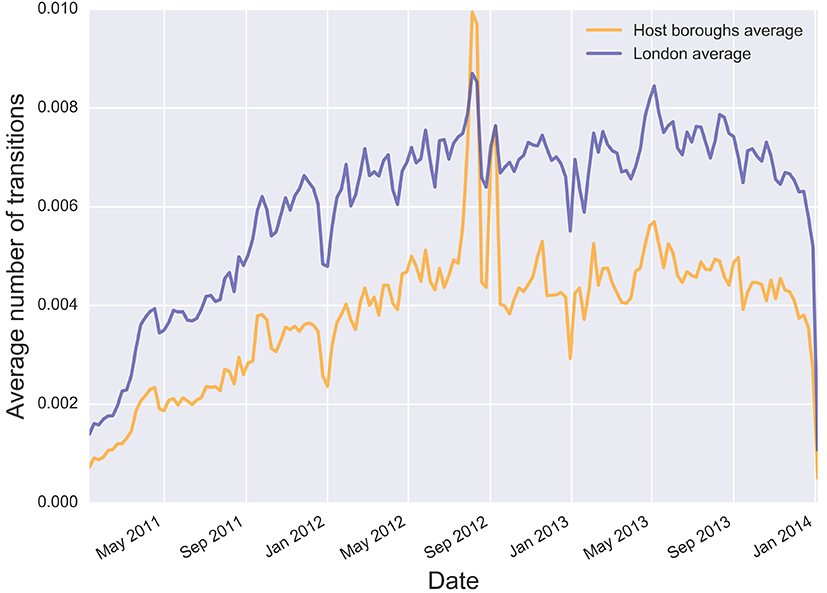}
	\caption{Average number of transition flows in host boroughs in total and London}
\end{figure}

\begin{figure}
	\centering
	\includegraphics[width=8cm]{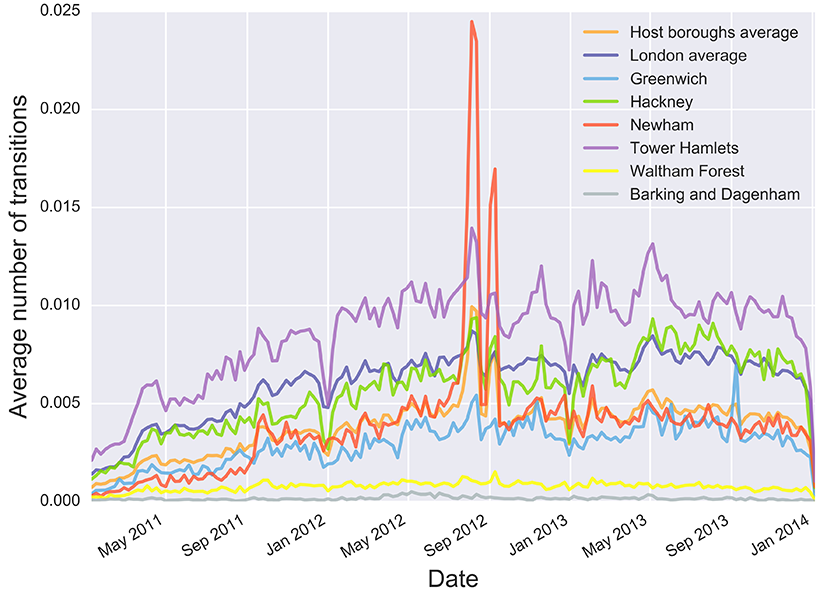}
	\caption{Average number of transition flows in each host borough and London }
\end{figure}

\subsubsection{Deprivation Changes}

As the SRF’s objectives clarified, the primary ambition of Olympic regeneration policy is to eliminate the radical deprivation in East London. To explore whether the deprivation level has changed before and after the Games, we compare IMD2010 and IMD2015 as well as their sub domains of deprivation in health, employment, education, housing, income, crime and living environment. 

Figure 6 presents the changes in borough ranks on average IMD and sub domain scores for London boroughs between 2010 and 2015, where the positive values mean the deprivation has been reduced while negative ones mean it became worse. From the figure we can tell that the overall IMD deprivation of East London has been improved in general. To be more precise, Waltham Forest, Newham, Hackney and Greenwich are the four host boroughs whose overall deprivation was reduced between 2010 and 2015. The most significant improvement happened in Greenwich, whose ranking went up by 6 positions from the 8th most deprived borough to 14th in London. However, Barking and Dagenham, and Tower Hamlets unfortunately became relatively more deprived in 2015 compared with 2010. 

When focusing on the changes in sub domains, it can be found that the deprivation in most domains has been improved. We determine the overall deprivation level reduced in the study area when the average of rank change for all host boroughs is positive. Under such a rule, we can see that income, employment, education, health, crime and living environment dimensions of deprivation have all improved. However, the deprivation in terms of Barriers to 
housing and services domain became greater within five of the six host boroughs. This outcome suggests that it has become harder for residents to find affordable housing in the Olympic boroughs after the Games. 

\begin{figure}
	\centering
	\includegraphics[width=8cm]{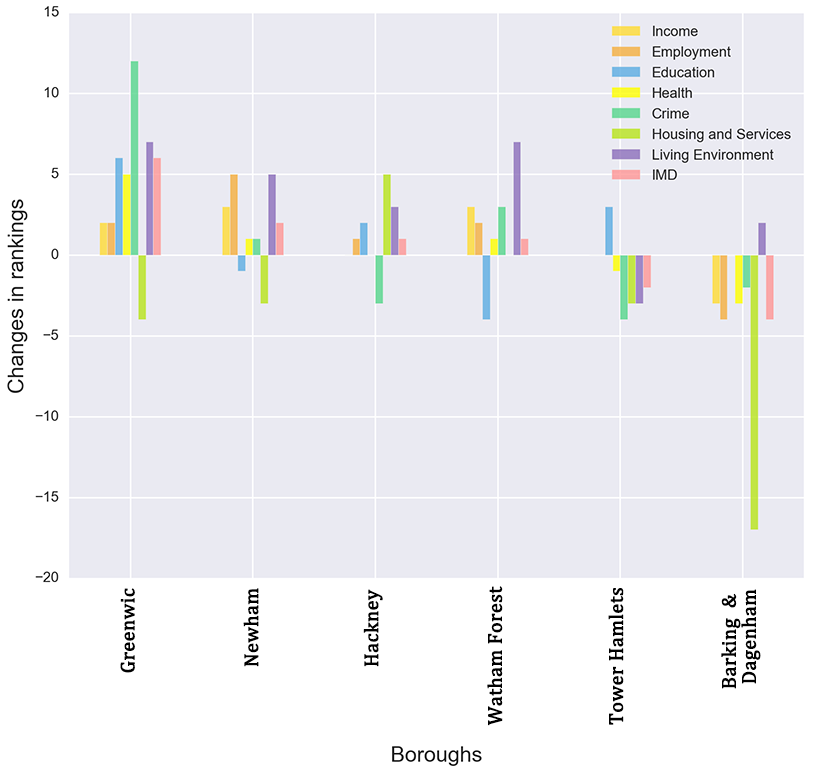}
	\caption{Changes in rankings of deprivation for host boroughs between 2010 and 2015}
\end{figure}

\section{Discussion and Conclusion }

Through the analyses above, some positive results are fortunately discovered, suggesting that regeneration indeed happened within East London to some degree. To evaluate the effectiveness of Olympic regeneration project by reference to the objectives in the SRF, we observe that the physical regeneration goal has been at least partly achieved as an increasing number of new businesses have been attracted to East London by a growth rate higher than London’s average.  The composition of these new venues is very diverse among which, food, nightlife spots, recreation, arts and entertainment categories are the main contributors. From a socio-economic perspective, the host boroughs of Waltham Forest, Newham, Hackney and Greenwich generally saw significant improvements in key indicators of deprivation, including income, employment, education, health, and living environment. Barriers to housing and services is the only deprivation indicator that became even worse after the Games. This finding indicates that the target of ‘home for all’ may fail to be met and reminds policy makers to pay more attention to the social problems coming along with the regeneration, such as the displacement of poorer local residents.  

When looking at the performance of each host borough, Newham, the main site of the Games, excels in venue density and transition flows growth during the Olympic period. In contrast, Barking and Dagenham, the youngest official host borough, who joined the Olympic project in April, 2011, performed worst in both physical and socioeconomic regeneration. This may be attributed to the fact that the study period in this research is too close to the time it became a member. Follow-up analysis would be needed to evaluate its long-term regeneration success as well as other boroughs within East London as a whole.

The datasets used and results concluded in this research are certainly not free of biases and limitations. The data on venues registered in Foursquare is not the absolute venue set in London, however, it does exhibit successfully patterns of change during the study period. It would also be improper to state that everything that is happening in East London is occurring because of the 2012 Olympics. However, it might be no exaggeration to say that regeneration did happen within the area of East London between 2010 and 2015, which coincides with the Olympic policy. The analysis presented in this work opens the door to using spatio-temporal big data to evaluate government’s regeneration policy from a dynamic view, and provides evidence and implications for further related policymaking processes. 

\section{Acknowledgements}
We acknowledge the support from the Cambridge Trust, the China Scholarship Council, and Foursquare.

\subsection{References} 
Cashman, R. (2009). Regenerating Sydney’s West: framing and adapting an Olympic vision, in Poynter, G. and MacRury, I. (eds.) Olympic Cities: 2012 and the Remaking of London. Farnham: Ashgate, 130-143. 
\smallskip \noindent \\
DCLG (2015). London 2012 Olympics: regeneration legacy evaluation framework. 
\smallskip \noindent \\
Daggitt, M. L., Noulas, A, Shaw, B., \& Mascolo, C. (2015). Tracking urban activity growth globally with big location data. Royal Society open science, 3(4), 150688. 
\smallskip \noindent \\
Dmitry, S. (2012). A marathon not a sprint? Legacy lessons for London. Centre for Cities. 
\smallskip \noindent \\
Galina, G. (2011). The Olympics’ employment and skills legacy: a literature review. Work and Employment Research Unit, University of Greenwich. 
\smallskip \noindent \\
Gavin, P. (2012). Mega events and the urban economy: what can Olympic cities learn from each other? http://olympicstudies.uab.es/lectures/web/pdf/Poynter\_\\eng.pdf 
\smallskip \noindent \\
Karamshuk, D., Noulas, A., Scellato, S., Nicosia, V, \& Mascolo, C. (2013). Geo-spotting: mining online location-based services for optimal retail store placement. Proceedings of the 19th ACM SIGKDD international conference on Knowledge discovery and data mining, August 11-14, Chicago, Illinois, USA.  
\smallskip \noindent \\
Keddie, J. (2014). Negotiating urban change in gentrifying London: Experiences of long-term residents and early gentrifiers in Bermondsey (PhD thesis).  
\smallskip \noindent \\
Lees, l. \& Clare, M. (2015). Arts-led regeneration in the UK: the rhetoric and the evidence on urban social inclusion. European Urban and Regional Studies, 22(3), 242-260. 
\smallskip \noindent \\
Macrury, I. (2008). Re-thinking the legacy 2012: the Olympics as commodity and gift. Twenty-First Century Society, 3(3), 297-312 
\smallskip \noindent \\
NEF (2008). Fools gold. Community Links. 
\smallskip \noindent \\
ODA (2010). The Olympic Park: towards a 10 year landscape management \& maintenance plan.  
\smallskip \noindent \\
Sue, B. (2010). Literature review: Olympic venues-regeneration legacy London assembly. Oxford Brookes University. 
\smallskip \noindent \\
SRF (2009). Strategic regeneration framework: an Olympic legacy for the host boroughs. 
\smallskip \noindent \\
Sassen, S. (2001). The Global City: New York, London, Tokyo, Princeton: University Press. 
\smallskip \noindent \\
Webber, R. (2007) The metropolitan habitus: its manifestations, locations and consumption profiles. Environment and Planning A, 39, 182-207.  
\smallskip \noindent \\
Ye, M., Vojnovic, I., \& Chen, G. (2015). The landscape of gentrification: exploring the diversity of “upgrading” processes in Hong Kong, 1986–2006. Urban Geography, 36(4), 471-503.

\end{document}